\documentclass[a4paper,11pt]{article}
\pdfoutput=1
\usepackage{graphicx}
\usepackage{fullpage}
\usepackage{jheppub} % for details on the use of the package, please
\usepackage{bbm}
\usepackage{amsfonts}
\usepackage{slashed}
\usepackage{subfigure}
\usepackage{array}
\usepackage{verbatim}
\usepackage{booktabs}
\usepackage{color, soul}
\usepackage{mathrsfs}
\usepackage{epsfig}
\usepackage{graphicx}% Include figure files
\usepackage{dcolumn}% Align table columns on decimal point
\usepackage{bm}% bold math
\usepackage{amsmath}
\usepackage{amssymb}
\usepackage{multirow}
\usepackage{caption}

\setstcolor{red}
\usepackage{makecell}

\title{\sffamily Correlation between $R_{D^{(*)}}$ and top quark FCNC decays in leptoquark models}

\author[a]{Tae Jeong Kim,}
\author[b,c]{Pyungwon Ko,}
\author[b,d]{Jinmian Li,}
\author[a]{Jiwon Park,}
\author[b]{Peiwen Wu}

\affiliation[a]{Hanyang University, Department of Physics Wangsimniro 222, Seoul 04763, Republic of Korea}
\affiliation[b]{School of Physics, KIAS, 85 Hoegiro, Seoul 02455, Republic of  Korea}
\affiliation[c]{Quantum Universe Center, KIAS, 85 Hoegiro, Seoul 02455, Republic of Korea}
\affiliation[d]{College of Physical Science and Technology, Sichuan University, Chengdu, Sichuan 610065, China}

\emailAdd{taekim@hanyang.ac.kr}
\emailAdd{pko@kias.re.kr}
\emailAdd{phyljm@gmail.com}
\emailAdd{jiwon.park@cern.ch}
\emailAdd{pwwu@kias.re.kr}

\abstract{Some interpretations of $R_{D^{(*)}}$ anomaly in $B$ meson decay using leptoquark (LQ) models can also generate top quark decays through Flavor Changing Neutral Current (FCNC). In this work we focus on two LQs, i.e. scalar $S_1$ and vector $U_1$ which are both singlet under the $SU(2)_L$ gauge group in the Standard Model (SM). We investigate their implications on the 3-body top FCNC decays $t\to c \ell_i \ell_j$ at tree level and the 2-body $t\to c V$ at one-loop level, with $\ell$ being the SM leptons and $V=\gamma, Z, g$ being the SM gauge bosons. We utilize the $2\sigma$ parameter fitting ranges of the LQ models and find that $Br(t\to c \ell_i \ell_j)$ at tree level can reach $\mathcal{O}(10^{-6})$ and $Br(t\to c V)$ at one-loop level can reach $\mathcal{O}(10^{-10})$. Some quick collider search prospects are also analyzed.}

\begin{document}
\maketitle \indent
\newpage

\section{\label{sec-introduction}Introduction}

The deviations of $B$ meson decays from the Standard Model (SM) predictions have attracted a lot of attention in the past several years \cite{Lees:2012xj,Lees:2013uzd,Huschle:2015rga,Aaij:2015yra,Hirose:2016wfn,Sato:2016svk,Abdesselam:2016cgx}. Two significant processes are $R_{D^{(*)}}$ and $R_{K^{(*)}}$ which are defined through ratios of Branching Ratios (BRs) as follows:
\begin{eqnarray}
R_{D^{(*)}} = \left. \frac{Br(B\to D^{(*)} \tau\bar{\nu})}{Br(B\to D^{(*)} l \bar{\nu})}\right |_{l\in \{e,\mu\}}, \quad R_{K^{(*)}}^{[q_1^2, q_2^2]} =  \frac{\int^{q_2^2}_{q_1^2}dq^2 \frac{\partial}{\partial q^2}Br(B\to K^{(*)} \mu\mu)}{\int^{q_2^2}_{q_1^2}dq^2 \frac{\partial}{\partial q^2}Br(B\to K^{(*)} ee)}  \,,
\label{eq:RD-RK}
\end{eqnarray}
with $q^2=(p_{l^+}+p_{l^-})^2$ between $q_1^2$ and $q_2^2$ in units of ${\text{GeV}}^2$. For $R_{D^{(*)}}$, the world-averaged results after the recent update from Belle Collaboration \cite{Abdesselam:2019dgh} are\footnote{These updates do not change significantly from the previous ones when considering uncertainties, i.e. $R_{D} =0.407 \pm 0.046$ and $R_{D^{* }} = 0.306 \pm 0.015$ \cite{Amhis:2016xyh}. Our main conclusions  in this work are not affected much, especially for the order of magnitude in our numerical results.}:
\begin{eqnarray}
R_{D} =0.334 \pm 0.031 \,,\qquad R_{D^{* }} = 0.297 \pm 0.015 \,, \\
R_{D}^\mathrm{SM}=0.299 \pm 0.003 \,,\qquad R_{D^{*}}^\mathrm{SM}=0.258 \pm 0.005,
\end{eqnarray}
which are larger than the SM predictions at about $3.1\sigma$ \cite{Lattice:2015rga,Na:2015kha,Aoki:2016frl,Bigi:2017jbd,Bernlochner:2017jka,Jaiswal:2017rve}.

The latest measurements of $R_{K}$ at LHCb \cite{Aaij:2019wad} and $R_{K^{*}}$ at Belle \cite{Abdesselam:2019wac} are\footnote{$R_{K^*}^{[1.1,6]}$ cited are using the combined charged and neutral channels in Belle's measurement. $R_{K^{(*)}}$ in other energy bins can be found in \cite{Abdesselam:2019wac}.}:
\begin{eqnarray}
R_K^{[1.1,6]}=0.846^{+ 0.060}_{-0.054}({\rm stat.})^{+0.016}_{-0.014}({\rm sys.}), \,\, R_{K^*}^{[1.1,6]}=0.96^{+0.45}_{-0.29}({\rm stat.})^{+0.11}_{-0.11}({\rm sys.}),
\end{eqnarray}
which are smaller than the SM predictions shown below \cite{Bordone:2016gaq,Hiller:2003js} at around $2\sigma$.
\begin{eqnarray}
R_{K}^{[1.1,6],\mathrm{SM}}=1.00\pm 0.01, \,\, R_{K^{*}}^{[1.1,6],\mathrm{SM}}=1.00\pm 0.01.
\end{eqnarray}

The observed deviations of $R_{D^{(\ast)}}^\mathrm{exp}> R_{D^{(\ast)}}^\mathrm{SM}$ and $R_{K^{(\ast)}}^\mathrm{exp}< R_{K^{(\ast)}}^\mathrm{SM}$ have motivated many interpretations by imposing physics beyond the SM (see a recent review in \cite{Bifani:2018zmi} and references therein). Many of the theoretical proposals introduce additional charged scalars 
\cite{Crivellin:2012ye,Tanaka:2012nw,Celis:2012dk,Crivellin:2013wna,Crivellin:2015hha,Chen:2017eby,Iguro:2017ysu,Chen:2018hqy,Ko:2017lzd,Ko:2012sv,Cheung:2016fjo} 
and/or vectors
\cite{Greljo:2015mma,Boucenna:2016wpr,Boucenna:2016qad,Megias:2017ove,Alonso:2017uky,Alonso:2017bff,Duan:2018akc,Bian:2017xzg,Bian:2017rpg,Ko:2017yrd} to mediate the Charged Current (CC) in $R_{D^{(\ast)}}$ and Neutral Current (NC) in $R_{K^{(\ast)}}$, which can be realized in various UV-complete models. Recent discussions can be found in \cite{Biswas:2018snp,Blanke:2018yud,Biswas:2018iak,Iguro:2018vqb,Aebischer:2018acj,Dobrich:2018jyi,Faisel:2018bvs,Bansal:2018nwp,Kaneko:2018mcr,Mandal:2018kau,Maji:2018gvz,Babu:2018vrl,Alonso:2018vwa,Kamali:2018bdp,Faroughy:2018cyl,Roy:2018nwc,Geng:2018xzd,Hutauruk:2019crc,Baek:2019qte} and also recently in \cite{Aaij:2019okb,Popov:2019tyc,Davighi:2019jwf,Kamada:2019gpp,Hryczuk:2019nql,Trifinopoulos:2019lyo,Allanach:2019mfl,Cornella:2019hct,DelleRose:2019ukt,Kowalska:2019ley,Gherardi:2019zil,Shaw:2019fin,Aebischer:2019mlg,Crivellin:2019dun,Datta:2019zca,Ciuchini:2019usw,Shi:2019gxi,Blanke:2019qrx,Crivellin:2019qnh,Heeck:2019nmh,Asadi:2019xrc,Cali:2019nwp,Yan:2019hpm,Murgui:2019czp,Hou:2019uxa,Choudhury:2019ucz,Kumbhakar:2019njm,Datta:2019tuj,Zhang:2019jwp,Arbey:2019duh,Azizi:2019aaf,Arnan:2019uhr,Kumar:2019qbv,Bardhan:2019ljo,Alguero:2019ptt,Alok:2019ufo,Alok:2019uqc}.

In this work we are not going to be ambitious to explain both deviations, but limit ourselves to $R_{D^{(\ast)}}$ interpretations in the leptoquark (LQ) models \cite{Fajfer:2012jt,Deshpande:2012rr,Sakaki:2013bfa,Alonso:2015sja,Calibbi:2015kma,Bauer:2015knc,Fajfer:2015ycq,Barbieri:2015yvd,Deshpand:2016cpw,Li:2016vvp,Sahoo:2016pet,Becirevic:2016yqi,Dumont:2016xpj,Das:2016vkr,Barbieri:2016las,Chen:2017hir,Altmannshofer:2017poe,Alok:2017sui,Alok:2017jgr,Alok:2017jaf,Crivellin:2018yvo} and its interesting correlations to the top quark Flavor Changing Neutral Current (FCNC) decays. Recently, several studies investigated the implications of the six types of LQ models on $R_{D^{(\ast)}}$ and $R_{K^{(\ast)}}$, including three scalars $\{ S_1, R_2, S_3 \}$ and three vectors $\{ U_1, V_2, U_3 \}$ where the subscript denotes $2 T_3 +1$ with $T_3$ being the LQ's weak isospin. Results show that three of them are still capable of accommodating $R_{D^{(\ast)}}$ excess while satisfying other flavor constraints, i.e. SM $SU(2)_L$ singlet scalar  $S_1$ and vector  $U_1$, as well as $SU(2)_L$ doublet scalar  $R_2$.

In this work we  concentrate on the two $SU(2)_L$ singlet scenarios, i.e. $S_1$ and $U_1$, motivated by the simplicity and, as we will see later, the resulting clear correlation patterns between $R_{D^{(\ast)}}$ explanations and the top decays through FCNC. Note that the benchmark parameters we utilize in the numerical analysis may not be able to produce the observed $R_{K^{(\ast)}}$ anomaly. For example, requiring $S_1$ to explain $R_{K^{(\ast)}}$ appears to result in conflict with $R_D^{\mu/e}=Br(B\to D\mu \nu)/Br(B\to D e\nu)$ \cite{Becirevic:2016oho}. On the contrary, it has been shown that $U_1$ can still simultaneously generate the observed $R_{D^{(\ast)}}$ and $R_{K^{(\ast)}}$ \cite{Angelescu:2018tyl}. Putting aside the complexities in accommodating both anomalies, in this work we will exclusively investigate the $R_{D^{(\ast)}}$ interpretation and the interesting correlations with the top quark FCNC when introducing LQ $S_1$ or $U_1$.

This paper is organized as follows. In Section \ref{sec-2} we briefly capture the Lagrangian we consider for the scalar LQ $S_1$ and the vector LQ $U_1$, and the effective operators they generate in low-energy processes for $R_{D^{(\ast)}}$. In Section \ref{sec-3} we present the results for top FCNC decays induced at both tree level and one-loop level. Collider search prospects are given in Section \ref{sec-collider} and Conclusion will be given in Section \ref{sec-conclusion}. Appendix includes full expressions of one-loop Wilson coefficients of $t\to cV$ at one-loop level induced by the scalar LQ $S_1$.

\section{\label{sec-2}LQ $S_1$ and $U_1$ for $R_{D^{(*)}}$}

In this section we briefly capture the low-energy theory in terms of effective operators for $R_{D^{(*)}}$ and the Wilson coefficients generated by  the scalar LQ  $S_1$ and vector LQ $U_1$, respectively. Then we present the theoretical correlations between $R_{D^{(*)}}$ and BRs of top FCNC. We denote LQ as $(SU(3)_c,SU(2)_L)_Y$ which is its representation in the SM gauge group \cite{Angelescu:2018tyl,Dorsner:2016wpm}. Considering the misalignments between gauge and mass eigenstates in the quark sector, we define the left-handed quark doublet as $Q_i=[(V^\dagger u_L)_i \,\,d_{L\,i}]^T$ where $V$ is the Cabibbo-Kobayashi-Maskawa (CKM) matrix.

As mentioned earlier, we will focus on two LQs which are both singlet under the SM $SU(2)_L$ group, i.e. scalar $S_1\equiv (\mathbf{\bar{3}},\mathbf{1})_{1/3}$ and vector $U_1=(\mathbf{3},\mathbf{1})_{2/3}$. Their interactions with the SM fields we consider are
\begin{eqnarray}
\mathcal{L}_{S_1} &= g_{1L}^{ij} \, \overline{Q^C_i} i\tau_2 L_j S_1\,  + g_{1R}^{ij} \,\overline{u^C_{R\,i}} e_{R\,j}S_1\,  +\mathrm{h.c.}, \\
\label{eq-L-S1}
\mathcal{L}_{U_1} &= h_{1L}^{ij} \, \overline{Q}_i \gamma_\mu L_j U_1^\mu + h_{1R}^{ij} \, \overline{d}_{R\,i} \gamma_\mu  \ell_{R\,j} U_1^\mu+\mathrm{h.c.},
\label{eq-L-U1}
\end{eqnarray}
where $g_{1L}^{ij},g_{1R}^{ij}$ and $h_{1L}^{ij},h_{1R}^{ij}$ are matrices of new Yukawa interactions in the general case, and $\tau_2$ is the second Pauli matrix. We have neglected the terms of diquark couplings to LQ to ensure the stability of proton \cite{Dorsner:2016wpm}. Note again that we have chosen the form of the left-handed quark doublet as $Q_i=[(V^\dagger u_L)_i \, \,d_{L\,i}]^T$ in which the down-type quarks are mass eigenstates. Therefore it will be $(Vg_{1L})^{ij}$ and $(Vh_{1L})^{ij}$ that enter the interactions involving up-type left handed quarks.

The general low-energy effective dimension-six operators involved in $B\to D^{(*)} \tau\bar{\nu}$ are \cite{Sakaki:2013bfa,Dumont:2016xpj}
\begin{equation}
   -  \mathcal{L}_\text{eff} =     ( C_\text{SM} \delta_{l \tau}    + C_{ V_1}^l)  O_{ V_1}^l + C_{ V_2}^l  O_{ V_2}^l  + C_{ S_1}^l  O_{ S_1}^l + C_{ S_2}^l  O_{ S_2}^l + C_{ T}^l  O_{ T}^l  \,,
\end{equation}
with $l=1,2,3$ being the neutrino generation index. $C_\text{SM} = 2 \sqrt 2 G_F V_{cb}$ is the SM contribution where $G_F$ is the Fermi constant. Operators above are defined as
\begin{align}
     &  O_{ V_1}^l = (\bar c_L \gamma^\mu b_L)(\bar \tau_L \gamma_\mu \nu_{lL}) \,, \quad
      O_{ V_2}^l = (\bar c_R \gamma^\mu b_R)(\bar \tau_L \gamma_\mu \nu_{lL}) \,, \\
     &  O_{ S_1}^l = (\bar c_L b_R)(\bar \tau_R \nu_{lL}) \,, \quad\quad\quad
       O_{ S_2}^l = (\bar c_R b_L)(\bar \tau_R \nu_{lL}) \,, \\
     &  O_{ T}^l = (\bar c_R \sigma^{\mu\nu} b_L)(\bar \tau_R \sigma_{\mu\nu} \nu_{lL}) \,.
\end{align}
The Wilson coefficients generated by $S_1$ and $U_1$ at the energy scale $\mu = M_{\text{LQ}}$ are
\begin{align}
      & C_{ V_1}^l =  \sum_{k=1}^3 V_{k3} 
      \left(       {g_{1L}^{kl}g_{1L}^{23*} \over 2M_{S_1}^2} + {h_{1L}^{2l}h_{1L}^{k3*} \over M_{U_1}^2}       \right) \,, \quad
       C_{ V_2}^l = 0 \,, \\
      & C_{ S_1}^l = \sum_{k=1}^3 V_{k3} 
      \left(       - {2h_{1L}^{2l}h_{1R}^{k3*} \over M_{U_1}^2}       \right) \,, \quad\quad\quad\quad
       C_{ S_2}^l = \sum_{k=1}^3 V_{k3} 
      \left(       -{g_{1L}^{kl}g_{1R}^{23*} \over 2M_{S_1}^2}       \right) \,, \\
      & C_{T}^l = \sum_{k=1}^3 V_{k3} 
      \left(       {g_{1L}^{kl}g_{1R}^{23*} \over 8M_{S_1}^2}       \right) \,.
\label{eq:WC-LQ}
\end{align}
For simplicity, in the following we only consider terms with $k=3$ and $V_{33}\approx 1$ for the LQ contributions\footnote{To make it consistent, in the following calculations we also ignored terms in LQ-quark-lepton couplings that are induced by non-diagonal CKM elements, i.e. from the up-type quark mixings defined in $Q_i=[(V^\dagger u_L)_i \, \,d_{L\,i}]^T$. We checked that the dropped terms are negligibly small.}. We note that \cite{Dumont:2016xpj} has provided the parameter ranges for various LQ models which can fit the $R_{D^{(*)}}$ data (see Table.II therein), as well as how they confront other flavor constraints. For example, a small $g_{1L}^{2l}$ can help $S_1$ pass the constraints from $\bar{B}\to X_s \nu\bar{\nu}$ while having available $g_{1L}^{3l}g_{1R}^{23*}$ to interpret $R_{D^{(*)}}$. Note that there are only two parameters in our analysis when choosing a certain generation index $l$, i.e. $g_{1L}^{3l}, g_{1R}^{23}$ for $S_1$ and $h_{1L}^{2l},h_{1L}^{k3}$ for $U_1$, which is different from the more complex textures in other works, e.g. \cite{Angelescu:2018tyl,Cai:2017wry}. Our choices can result in clear correlations between $R_{D^{(*)}}$ and top FCNC decays.

\section{\label{sec-3}LQ $S_1$ and $U_1$ for top quark FCNC}

\begin{figure}[ht]
\begin{center}
\includegraphics[width=6cm]{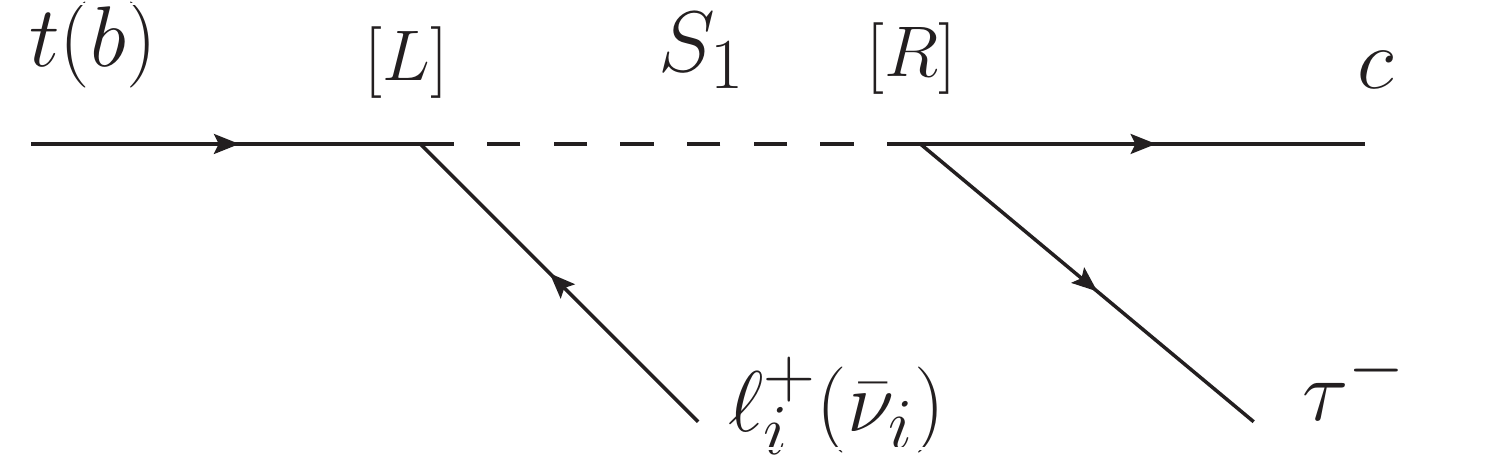}
\includegraphics[width=6cm]{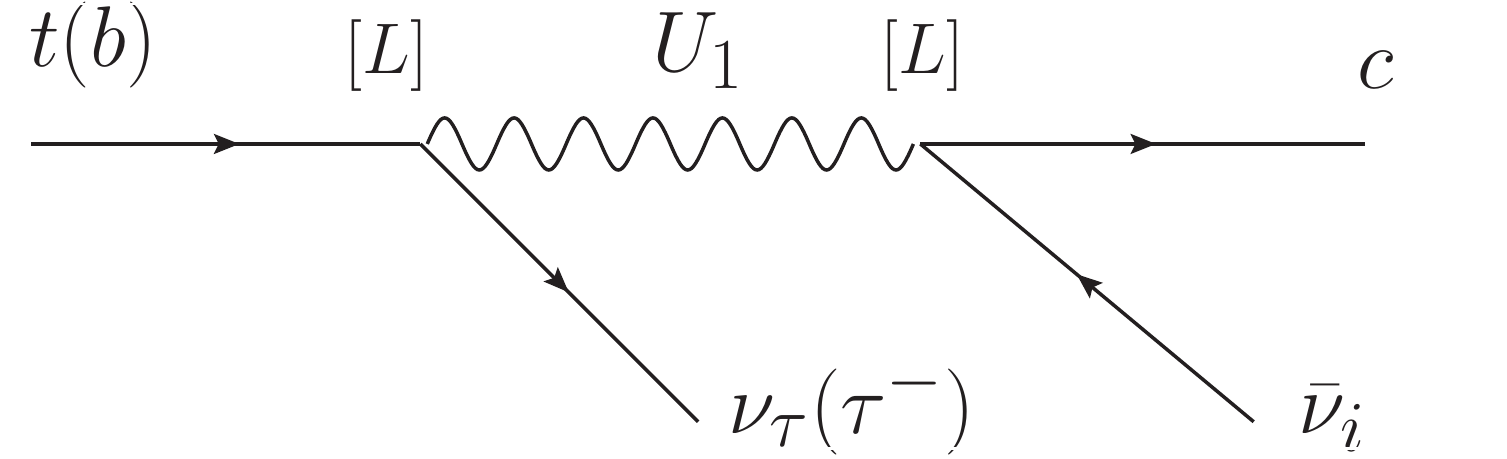}
\caption{Tree-level top FCNC decays considered in this work, induced by SU(2) singlet scalar LQ $S_1$  and vector LQ $U_1$.}
\label{fig:dia-topFCNC-tree}
\end{center}
\end{figure}

\begin{figure}[ht]
\begin{center}
\includegraphics[width=15cm]{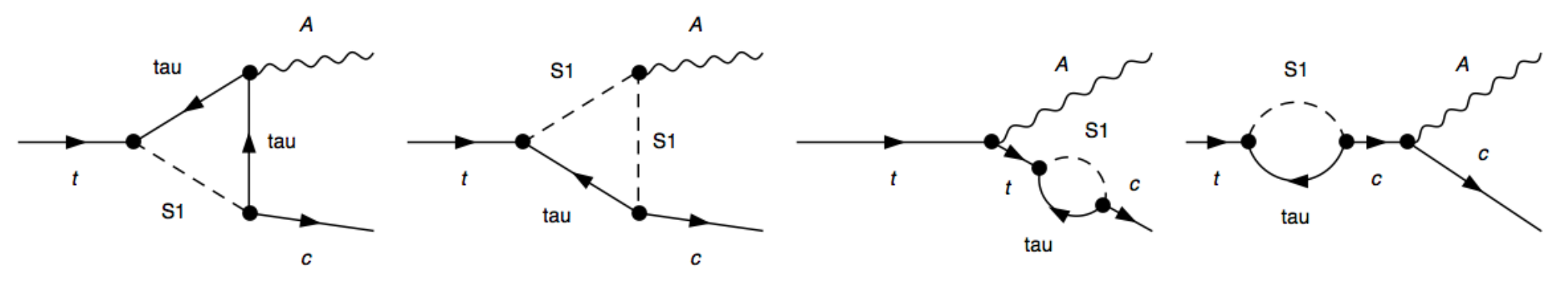}
\caption{One-loop top FCNC decays of $t\to c\gamma$ considered in this work, induced by SU(2) singlet scalar LQ $S_1$.}
\label{fig:dia-topFCNC-one-loop-S1}
\end{center}
\end{figure}

\begin{figure}[ht]
\begin{center}
\includegraphics[width=12cm]{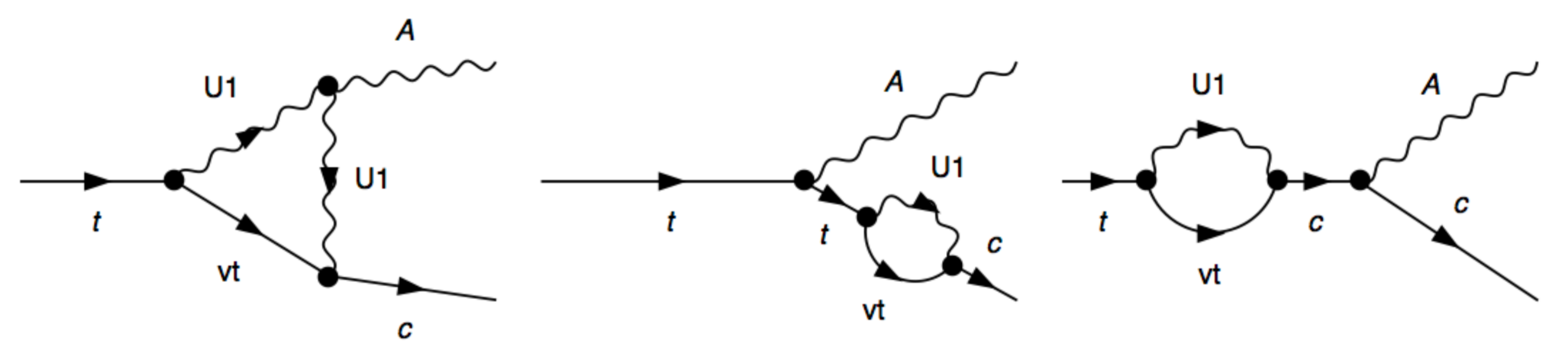}
\caption{One-loop top FCNC decays of $t\to c\gamma$ induced by SU(2) singlet vector LQ $U_1$. Note that we do not calculate these diagrams in this work, due to the lack of ultraviolet completion for vector LQ $U_1$ in our phenomenological studies. See more discussions in Section \ref{sec-one-loop}.}
\label{fig:dia-topFCNC-one-loop-U1}
\end{center}
\end{figure}

Diagrams of $S_1, U_1$ contributions to top FCNC at tree level $t \to c \tau^- \ell^+_i$ and $t \to c \nu_{\tau} \bar{\nu}_i$ are provided in Fig.\ref{fig:dia-topFCNC-tree} with $i$ denoting the lepton generation index. Square brackets indicate the chirality of couplings and replacement with particles in the round brackets generate processes involved in $R_{D^{(*)}}$. In Fig.\ref{fig:dia-topFCNC-one-loop-S1} and Fig.\ref{fig:dia-topFCNC-one-loop-U1} we also show the one-loop contributions to top FCNC $t \to c \gamma$ from $S_1$ and $U_1$, respectively, in which replacing external photon $\gamma$ with $Z$ boson or gluon $g$ with applicable vertices is straightforward.

In the numerical analysis, we utilize the parameter ranges in \cite{Dumont:2016xpj} for various LQ models which can fit the $R_{D^{(*)}}$ data at $2\sigma$ level (see Table.II therein). We remind ourselves that moderate differences in the $2\sigma$ ranges of parameters presented in different papers do not affect the order of magnitude in top FCNC BRs we will discuss. To be more clear, the parameter ranges we take from \cite{Dumont:2016xpj} in the numerical studies are summarized in Table.\ref{Table:parameters}. For simplicity, we assume all parameters are real in our analysis.
\begin{table}[t]
\begin{center}
\begin{tabular}{cc}
 \hline\hline       
   LQ     &  $2\sigma$ range for $\bar B \to D^{(*)} \tau\bar\nu$ \\
 \hline       
   $S_1$  & $g_{1L}^{3l} g_{1R}^{23*} \in \left( \frac{M_{S_1}}{\text{1 TeV}} \right)^2 \times \left\{ \begin{array}{ll}
   (1.64, 1.81)  & \quad l=1,2 \\ 
   (-0.87, -0.54) & \quad l=3
    \end{array} \right.$ \\
 \hline
   $U_1$  &  $h_{1L}^{2l}h_{1L}^{33*}  \in \left( \frac{M_{U_1}}{\text{1 TeV}} \right)^2 \times  \left\{ \begin{array}{ll}
 (0.52, 0.84) & \quad l=1,2 \\
 (-2.94,-2.80) & \quad l=3
  \end{array} \right.$ \\
\hline\hline
\end{tabular}
\caption{
Parameter ranges we utilize in numerical calculations, taken from Table.II of \cite{Dumont:2016xpj}. For simplicity, we assume all parameters are real in our analysis.}
\label{Table:parameters}
\end{center}
\end{table}

\subsection{\label{sec-tree}Tree level}

One important feature in the top FCNC decay induced at tree level by LQ $S_1$ and $U_1$ is that heavy LQ can be reasonably integrated out into effective coefficients in the amplitude, i.e. $\frac{ g_{1L} g_{1R} }{M_{S_1}^2}$ and $\frac{ h_{1L} h_{1R} }{M_{U_1}^2}$, which contribute as a whole piece in both the top FCNC decay and the $R_{D^{(*)}}$. This infers an interesting correlation between the two processes despite the specific values of the couplings and LQ masses, as long as LQ masses are heavy enough to justify the effective coefficients as good approximations of the full calculations.

The top FCNC BRs in Fig.\ref{fig:dia-topFCNC-tree} can be approximated as follows.
\begin{eqnarray}
S_1: Br(t \to c \tau^- \ell^+_l) &\approx&\frac{1}{\Gamma_{t,SM}} \Big( \frac{m_t^5}{6144 \pi^3} \Big) | \frac{ g^{3l}_{{1L}} g^{23*}_{{1R}} }{M_{S_1}^2} |^2 = 10^{-6} \times
\left\{ \begin{array}{ll}
 1.4 \sim 1.8 & \, l=1,2\\
 0.16 \sim 0.41 & \,  l=3
   \end{array} \right. \\
U_1:Br(t \to c \nu_{\tau} \bar{\nu}_l) &\approx&\frac{1}{\Gamma_{t,SM}} \Big( \frac{m_t^5}{1536 \pi^3} \Big) | \frac{ h^{33}_{{1L}} h^{2l*}_{{1L}} }{M_{U_1}^2} |^2 = 10^{-6} \times
\left\{ \begin{array}{ll}
 0.58 \sim 1.5 & \, l=1,2\\
 17 \sim 19 & \,  l=3
   \end{array} \right. 
\label{eq:NR-tree}
\end{eqnarray}
In the above, we take the SM parameters as $m_c\simeq m_\tau \simeq 0, m_t= 172 \, {\rm GeV}$ and $\Gamma_{t,SM}=1.5$ GeV, while $\frac{ g_{1L} g_{1R} }{M_{S_1}^2}$ and $\frac{ h_{1L} h_{1R} }{M_{U_1}^2}$ are taken from Table.\ref{Table:parameters}. The analytic expressions are approximations by integrating out LQ propagators, while the numerical results are obtained from full calculations using MadGraph \cite{Alwall:2014hca} with model files generated by FeynRules \cite{Alloul:2013bka}.

Note again that the connection between $R_{D^{(*)}}$ and top quark 3-body FCNC decays $Br(t \to c \ell_i\ell_j)\sim10^{-6}$ shown above do not depend directly on the specific values of couplings and LQ masses, but on the effective coefficients $\frac{ g_{1L} g_{1R} }{M_{S_1}^2}$ and $\frac{ h_{1L} h_{1R} }{M_{U_1}^2}$ as a whole piece. It holds well for sufficiently heavy $M_\text{LQ}$ ($\gtrsim 1$ TeV) which can justify the good approximations and suppress the high order terms $\propto \frac{m_t^2}{M^2_\text{LQ}}$ in the full calculation.

\subsection{\label{sec-one-loop}One-loop level}

For $t\to cV$ with $V=\gamma,g,Z$ at one-loop level, the amplitudes can be expressed in the following form:
\begin{equation}
i {\mathcal M}_{tcV}=\bar u(p_2)\, \Gamma^\mu_{tcV}  u(p_1)\, \epsilon_\mu(k,\lambda)~,
\label{eq:deltaM}
\end{equation}
where $p_1, p_2$, and $k$ denote the 4-momenta of the incoming top quark,
outgoing charm quark and the outgoing gauge boson, respectively, and
$\epsilon_\mu(k,\lambda)$ is the polarization vector of the outgoing gauge
boson. The vertices $\Gamma^\mu$ can be decomposed as follows \cite{Lopez1997-FCNC} when external particles are on-shell:
\begin{eqnarray}
\Gamma^\mu_{tcZ}&=&\gamma^\mu (P_L f^{Z}_{VL}+ P_R f^{Z}_{VR})
+ i\sigma^{\mu\nu}k_\nu (P_L f^{Z}_{TL} + P_R f^{Z}_{TR})\, ,
\label{eq:VtcZ}\\
\Gamma^\mu_{tc\gamma}&=&i\sigma^{\mu\nu}k_\nu (P_L f^{\gamma}_{TL} + P_R f^{\gamma}_{TR})\, ,
\label{eq:VtcA}\\
\Gamma^\mu_{tcg}&=& T^a i\sigma^{\mu\nu}k_\nu (P_L f^{g}_{TL} + P_R f^{g}_{TL})\,,
\label{eq:Vtcg}
\end{eqnarray}
with $P_{R,L}={1\over2}(1\pm\gamma_5)$, $\sigma^{\mu\nu}={i\over 2}[\gamma^\mu,\gamma^\nu]$ and $T^a$ are the $SU(3)$ color generators with $a=1,...,8$. The partial widths are
\begin{eqnarray}
\Gamma(t\rightarrow cZ)&=&{m_t^3 \over 32 \pi m_Z^2}(1- {m_Z^2\over m_t^2})^2
\Bigl[(1+2{m_Z^2\over m_t^2})(|f^{Z}_{VL}|^2+|f^{Z}_{VR}|^2) \nonumber\\
&&-6{m_Z^2 \over m_t}\,Re\Big (f^{Z}_{VL}f^{Z*}_{TR}+
f^{Z}_{TL} f^{Z*}_{VR}\Big) +m_Z^2(2+
{m_Z^2 \over m_t^2})(|f^{Z}_{TL}|^2+|f^{Z}_{TR}|^2)\Bigr]\, ,\\
\Gamma(t\rightarrow c\gamma)&=&{m_t^3\over 16 \pi } (|f^{\gamma}_{TL}|^2+|f^{\gamma}_{TR}|^2)\, ,\label{eq:BrVtcA}\\
\Gamma(t\rightarrow cg)&=&C_F\,{m_t^3\over 16 \pi } (|f^{g}_{TL}|^2+|f^{g}_{TR}|^2)\, ,\label{eq:BrVtcg}
\end{eqnarray}
where $C_F=(N_c^2-1)/2N_c$ with $N_c=3$ is the Casimir factor of $SU(N)$ and we set $m_c=0$ for simplicity. 
We use FeynArts/FormCalc \cite{Hahn:2000kx,Hahn:1998yk} to perform the one-loop calculations which is then linked to LoopTools \cite{Hahn:1998yk} to obtain numerical results. 

First of all, we note that in the case of vector LQ $U_1$, the model is non-renormalizable by introducing a single vector LQ $U_1$. This results in a divergent $U_1$ contribution to $t\to cV$ at one-loop level, unless the ultraviolet (UV) completion is established to generate the $U_1$ mass (see e.g. \cite{Assad:2017iib,DiLuzio:2017vat,Bordone:2017bld,Calibbi:2017qbu,Blanke:2018sro,Barbieri:2017tuq,Greljo:2018tuh}). The approach will be model-dependent and we will not address it further in this work. More discussions on effects of $U_1$ at one-loop level can be found in, e.g. \cite{Crivellin:2018yvo}. 

In the full calculations we include both the SM and the LQ contributions to take into account the interference effects. For the LQ $S_1$ contributions, we present the full expressions of Wilson coefficients at one-loop level in the Appendix. In the heavy mass range $M_{S_1}\simeq1\, {\rm TeV}$ which indicates $M_{S_1}\gg m_t,m_c,m_\tau$, one can have approximated results, especially for the massless gauge bosons $V=\gamma, g$. By setting $m_c=0$ and taking $x_\tau=m_\tau^2/M_{S_1}^2, x_t=m_t^2/M_{S_1}^2$, we have:
\begin{eqnarray}
f^{g}_{TL} &\simeq& \frac{1  }{16 \pi ^2} g_s m_{\tau }  \frac{g_{1 L}^{33} g_{1 R}^{23*} }{M_{S_1}^2} \nonumber \\
& &\times \frac{1}{12} \big(  -6 \left( 22 x_{\tau } x_t +3 x_t+16 x_{\tau }+6\right) \log x_{\tau }-49 x_t-48 \big) , \label{eq:fVtcg} \\
f^{\gamma}_{TL} &\simeq& - \frac{1  }{16 \pi ^2} e m_{\tau }  \frac{g_{1 L}^{33} g_{1 R}^{23*} }{M_{S_1}^2} \nonumber \\
& &\times \frac{1}{3}\big( \frac{1}{6} \left(14 x_t x_{\tau }+x_t+9 x_{\tau }+3\right)+\left(x_t+1\right) x_{\tau } \log x_{\tau }\big), \label{eq:fVtcA}\\
f^{g}_{TR} &=& f^{\gamma}_{TR}  = f^{Z}_{TR} = 0,
\end{eqnarray} 
where $g_s$ is the coupling of strong interaction. In the above, we have utilized Package-X \cite{Patel:2015tea,Patel:2016fam} to perform the loop function reductions. Note that the absence of right-handed dipole current is because of the coupling textures we considered in Table.\ref{Table:parameters}. In the case of massive $Z$ boson, the loop function approximations are tediously long \cite{Denner:1991kt} and we keep the full expressions in the Appendix.

In Fig.\ref{fig:BrtcV-loop-S1} we show the numerical results of $Br(t\to cV)$ with colors of red, green, blue indicating $V=\gamma, g, Z$, respectively. Solid lines include both the SM and the LQ contribution, while dashed lines are the SM predictions with the CKM matrix values taken from Particle Data Group \cite{PhysRevD.98.030001}.
In the left panel we choose $g_{1 L}^{33} g_{1 R}^{23*} = 1$ as an ordinary coupling benchmark to show the decoupling behavior of LQ $S_1$ contribution with respect to (w.r.t.) $M_{S_1}$. We see that when including the LQ $S_1$ contributions with $M_{S_1}\simeq 1 \,{\rm TeV}$, $Br(t\to c\gamma)$ ($Br(t\to cZ)$) are increased by a factor of about 2000 (400) from the SM predictions $5\times 10^{-14}$ ($1\times 10^{-14}$) to around $1\times 10^{-10}$ ($4\times 10^{-12}$). However, there is only a mild enhancement by a factor of around 3 for $Br(t\to cg)$, which is from $6\times 10^{-12}$ in the SM to around $2\times 10^{-11}$ when including the $S_1$ contributions. However, with sufficiently heavy $M_{S_1}$ all values of $Br(t\to cV)$ will reduce to the SM predictions.

In the right panel, we fix $\frac{g_{1 L}^{33} g_{1 R}^{23*} }{M_{S_1}^2} =0.87$ which is the upper bound value of numerical fitting for LQ models to explain $R_{D^{(*)}}$ at $2\sigma$ (see Table.\ref{Table:parameters}).
To keep $g_{1 L}^{33}, g_{1 R}^{23*}$ perturbative in this set-up, $M_{S_1}$ should not be too heavy. In the region of $M_{S_1}\simeq \mathcal{O}(1) \,{\rm TeV}$, one can learn from Eq.(\ref{eq:fVtcg}) and Eq.(\ref{eq:fVtcA}) that we have small but almost stable $x_\tau, x_t \ll \mathcal{O}(1)$. When combined with the SM contributions, $Br(t\to cV)$ are also fairly stable for $M_{S_1}\gtrsim 2 \,{\rm TeV}$ with values around $1\times 10^{-10}, 1\times 10^{-11}, 5\times 10^{-13}$ for $V=\gamma, g, Z$, respectively.

\begin{figure}[ht]
\begin{center}
\includegraphics[width=7cm]{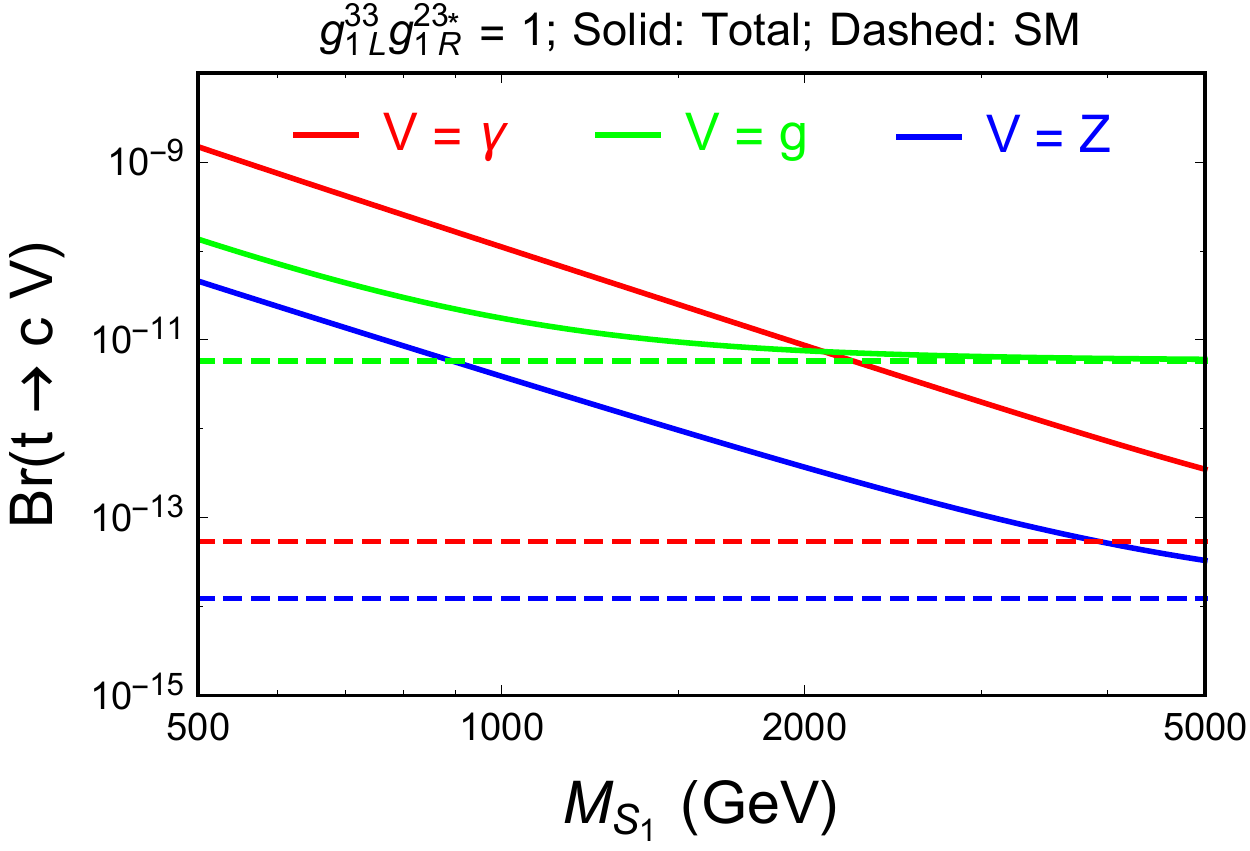}
\includegraphics[width=7cm]{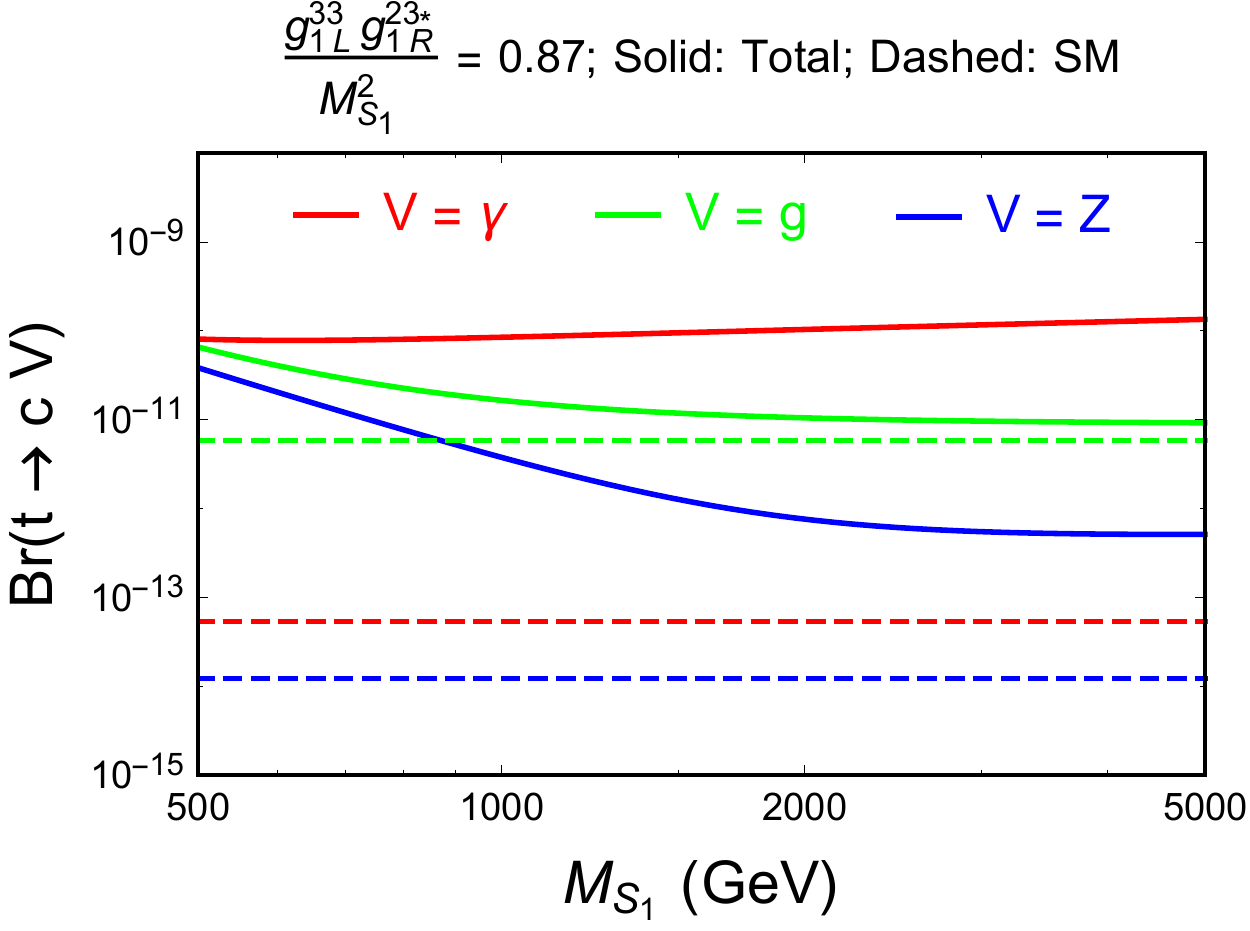}
\caption{$Br(t\to cV), \, V=\gamma,g,Z$ at one-loop level induced by SU(2) singlet scalar LQ $S_1$. In the left panel we choose $g_{1 L}^{33} g_{1 R}^{23*} = 1$ as an ordinary coupling benchmark to show the decoupling behavior of LQ $S_1$ contribution with respect to $M_{S_1}$. In the right panel, we fix $\frac{g_{1 L}^{33} g_{1 R}^{23*} }{M_{S_1}^2} =0.87$ which is the upper bound value of numerical fitting for LQ models to explain $R_{D^{(*)}}$ at $2\sigma$ (see Table.\ref{Table:parameters}). Solid lines include both the SM and the LQ contribution, while dashed lines are the SM predictions with the CKM matrix values taken from Particle Data Group \cite{PhysRevD.98.030001}.}
\label{fig:BrtcV-loop-S1}
\end{center}
\end{figure}

\section{\label{sec-collider}Collider search prospects}

The 2-body top quark FCNC decays have been searched intensively at the LHC. The current constraints on the Br$(t\to c \gamma)$, Br$(t\to c g)$ and Br$(t\to c Z)$ are found to be $2\times 10^{-3}$~\cite{Khachatryan:2015att}, $2\times 10^{-4}$~\cite{Aad:2015gea,Khachatryan:2016sib} and $2\times 10^{-4}$~\cite{CMS-PAS-TOP-17-017,Aaboud:2018nyl}, respectively. These are about six orders of magnitude above the predicted BR values $\mathcal{O}(10^{-10})$ at one-loop level induced by our LQ scenarios of explaining $R_{D^{(*)}}$, as presented in Fig.~\ref{fig:BrtcV-loop-S1}. Therefore, with such small BRs there is basically no hope to detect the signals of the 2-body top quark FCNC decays induced by $S_1$ explanation of $R_{D^{(*)}}$.

As for the 3-body top quark FCNC decays at tree level, our discussions in Sec.~\ref{sec-tree} show that the LQ explanation of the $R_{D^{(\ast)}}$ can induce $t \to c \mu \tau$, $t\to c\tau \tau$ and $t \to c \nu \nu$ with BRs $\sim10^{-6}$. In the following we perform some assessments of the search prospects for the 3-body top FCNC decays at the future upgraded LHC with integrated luminosity of 3000 fb$^{-1}$ and collision energy at 13 TeV. We will first consider the cut-and-count analysis which turns out to be not effective, then we proceed with further studies using multi-variate analysis techniques of Boosted Decision Tree (BDT) method.

\subsection{Cut-and-count analysis}
\label{sec-collider-1}

The LHC is a top quark factory. With integrated luminosity of 3000 fb$^{-1}$ and collision energy at 13 TeV, about $2.5 \times 10^{9}$ top quark pair events will be produced. If the LQ explanation of the $R_{D^{(\ast)}}$ can induce $t \to c \mu \tau$, $t\to c\tau \tau$ and $t \to c \nu \nu$ with BRs $\sim10^{-6}$, there will be $\sim 2500$ events which include at least one top quark decaying in these 3-body FCNC modes.
In order to suppress the multi-jet events and trigger the signal events, we require the other top quark in the top quark pair event to decay leptonically ($t \to b W, ~W \to \ell \nu$). This requirement still gives $\sim 500$ 3-body top quark FCNC events induced by our LQ scenarios of explaining $R_{D^{(*)}}$. The dominant SM backgrounds are $t\bar{t}$ with both top quarks decay through $t\to b W$, diboson production (VV), Drell-Yan process (DY) and $W$+jets events.

The following preselections will be applied to pick out the final state for each one of $t \to c \mu \tau$, $t\to c\tau \tau$ and $t \to c \nu \nu$.
\begin{itemize}
\item {Selection 1}: Exactly one lepton, at least three jets including exactly one $b$ jet and two $\tau$ jets.
\item {Selection 2}: Exactly two leptons, at least one muon, at least 2 jets including exactly one $b$ jet and one $\tau$ jet. 
\item {Selection 3}: Exactly one lepton and more than two jets in the final state, where one of the jet is $b$-tagged, the missing transverse energy $E_T^{\text{miss}}>80$ GeV.
\end{itemize}
In the event selections, our requirements include:
\begin{itemize}
\item The muons are required to have $p_T>30 (25)$ GeV and $|\eta|<2.4$, while the electrons should have $p_T>35 (30)$ GeV and $|\eta|<2.4$ for leading (sub-leading) lepton.
\item For both electrons and muons, the scalar sum of transverse momenta $H_T$ of all particles with $p_T > 0.5$ GeV that lie within a cone of radius $R = 0.3 (0.4)$ around the $e(\mu)$ should be less than 25(15)\% of the transverse momentum of $e(\mu)$.
\item The jets are reconstructed through anti-$k_T$ algorithm with radius parameter $R = 0.4$. Each should have $p_T>30$ GeV and $|\eta|<2.4$.
\item The $\tau$ jet tagging efficiency is 60\% with a QCD jet mis-identification rate of 1\% (5\%) for $p_T<40$ GeV ($p_T>40$ GeV). 
We set the $b$-tagging efficiency to be 68 \%, and the corresponding mis-tagging rates for the charm and light flavor jets are 0.12 and 0.01~\cite{Sirunyan:2017ezt}. 
\end{itemize}

The number of signal and background events after selections at HL-LHC are provided in Tab.~\ref{Table:selection}, in which the $W$+jet events in Selection 1 and 2 are negligible because of low statistics after the requirement of $b$ and $\tau$ jets. We can see that after the preliminary selection, the $t\bar{t}$ events are the dominant SM background and they are about $10^4\sim10^5$ times larger than signal events. We also found that with this simple cut-and-count method constructed above, the signal BRs above $\sim 5 \times 10^{-5}$ can be excluded at the $95\%$ confidence level (2 standard deviations). Therefore, the simple cut-and-count analysis is not powerful enough for probing the 3-body top quark FCNC decay signals with BR $\sim10^{-6}$ induced by our LQ scenarios of explaining $R_{D^{(*)}}$.

\begin{table}[htb]
\begin{center}
\begin{tabular}{c|c|c|c|c|c|c|c}
 \hline\hline       
  &  VV & DY  & $W$+jet & $t\bar{t}$ & $t\to c \mu\tau$ & $t \to c \tau \tau$ & $t \to c \nu \nu$ \\
 Selection 1 	& 9559		& 108095 	& - 			& 1189719 	& 28 		& 19  	& 0.3 \\
 Selection 2 	& 5433		& 54047 	& - 			& 839651 		& 39 		& 5 		& 0.0 \\
 Selection 3 	& 296814		& 594522 & 16530371 	& 64764862 	& 140 	& 94 		& 102\\  
  \hline\hline
\end{tabular}
\caption{The number of signal and background events after selections at 13 TeV LHC with integrated luminosity of 3000 fb$^{-1}$. The signals are the $t\bar{t}$ events with one top quark decaying leptonically and the other one decaying though $t \to c \mu \tau$, $t\to c\tau \tau$ and $t \to c \nu \nu$ induced by our LQ scenarios of explaining $R_{D^{(*)}}$ with a universal benchmark BR $\sim10^{-6}$. The $W$+jet events in Selection 1 and 2 are negligible because of low statistics after the requirement of $b$ and $\tau$ jets.}
\label{Table:selection}
\end{center}
\end{table}

\subsection{Multi-variate analysis}
\label{sec-collider-2}

\begin{figure}[ht]
\begin{center}
\includegraphics[width=10cm]{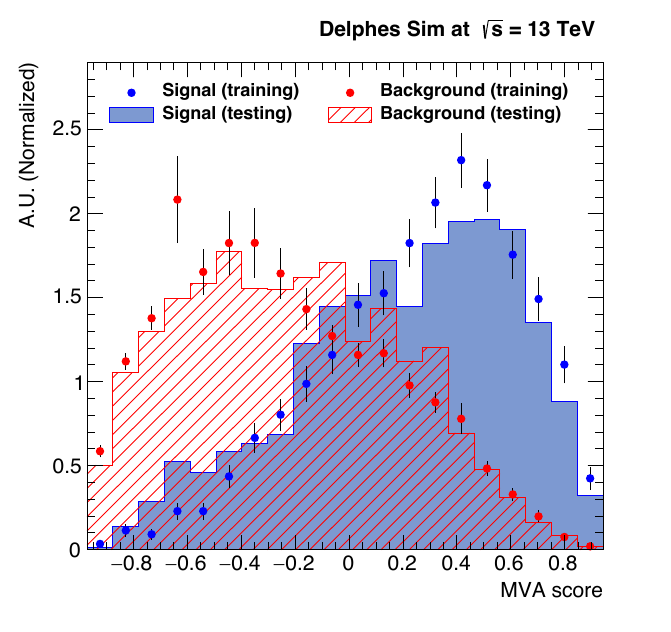}
\caption{Normalized BDT output distribution for signal and background events. The signal events tend to be close to 1 and the background events are close to -1. The training and testing samples are indicated by the dot and the filled histograms, respectively. The training and testing samples are compatible with each other which means there is no overfitting. We have utilized Delphes \cite{Ovyn:2009tx,deFavereau:2013fsa} to simulate the detector responses.}
\label{fig:fig-BDT}
\end{center}
\end{figure}

We proceed with some further studies using multi-variate analysis of BDT method, which is one of the machine learning techniques with the kinematic variables of the final objects, the angle distributions between leptons and $E_T^{\text{miss}}$, and so on. In our results shown in Fig.\ref{fig:fig-BDT}, the input variables we consider include:

\begin{itemize}
\item	multiplicity of jets, $b$ jet, $c$ jet and $\tau$ jet 
\item	$p_T, E_T^{\text{miss}}$ of the leading lepton 
\item	$p_T$ of the leading $\tau$ jet
\item	$H_T$ which includes jets, leptons and $E_T^{\text{miss}}$
\item	$p_T$ of leptons + $E_T^{\text{miss}}$, $p_T$ of leptons + $\tau$ jet
\item	$\Delta R(\tau, \ell), \Delta \phi(\ell, E_T^{\text{miss}}), \Delta \phi(\tau, E_T^{\text{miss}})$
\item	$\Delta \phi(\tau+\ell, E_T^{\text{miss}})$
\item	$\Delta R$($\ell$, leading $b$ jet)
\end{itemize}

In order to have a smooth distribution of the output BDT, we need to loosen the event selections. As an example, if we loosen the event Selection 1 to 1 lepton and 2 jets (one $b$ jet and one $\tau$ jet) for the signal of the $t \rightarrow c\mu\tau$ process, the signal significance defined by\footnote{Systematic uncertainties are not included in this quick multi-variate analysis given the limited statistics.} $S/\sqrt{B}$ goes down to 0.0138. In Fig.\ref{fig:fig-BDT} when focusing on the last 10 bins of signal and background in the MVA score distribution between $[0,1]$, we found that using BDT can help the cut-and-count method give an increased significance of 0.0185. Therefore, using the shape of the final BDT distribution we are able to increase the significance by $\mathcal{O}(30\%)$ compared to the cut-and-count method. However, to further increase the sensitivity in a more comprehensive analysis, we need to perform the shape analysis with more sophisticated statistical tools. As this involves dedicated data analysis with much more statistics, we would leave the technical improvement in future works.

\section{\label{sec-conclusion}Conclusion}

In this work we studied the correlation between the interpretations of $R_{D^{(*)}}$ anomaly in $B$ meson decay using LQ models and the top quark FCNC decays, i.e. 3-body processes $t\to c \ell_i \ell_j$ at tree level and 2-body processes $t\to c V$ at one-loop level, with $\ell$ being the SM leptons and $V=\gamma, Z, g$ being the SM gauge bosons. We focus on the scalar LQ $S_1$ and vector LQ $U_1$ which are both singlet under the SM $SU(2)_L$ gauge group. Utilizing the $2\sigma$ parameter fitting ranges of the LQ models, we find that 3-body processes $Br(t\to c \ell_i \ell_j)$ at tree level can reach $\mathcal{O}(10^{-6})$, and the 2-body processes $Br(t\to c V)$ induced by scalar LQ $S_1$ at one-loop level can reach $\mathcal{O}(10^{-10})$. We also provided quick estimations of the collider search prospects for a benchmark scenario $Br(t\to c \ell_i \ell_j)\sim 10^{-6}$ at 13 TeV HL-LHC with integrated luminosity of 3000 fb$^{-1}$. We found that the simple cut-and-count method can not give promising collider signal significance, but multi-variate analysis using BDT technique can provide reasonable improvements. More refined collider analyses are desirable which are left for future dedicated works.

\section*{Acknowledgement}
PW would like to thank Xiaofei Guo and Florian Staub for helpful discussions.
The work of P. Ko is supported by National Research Foundation of Korea Grant NRF-2019R1A2C3005009 (PK).
The work of T. J. Kim and J. Park was supported by Basic Science Research Program through 
the National Research Foundation of Korea (NRF) funded by the Ministry of Education, Grant No. NRF-2017R1A2B4002498
and the research fund of Hanyang University (HY-2015).

\section*{Appendix}

Here we present the Wilson coefficients of 2-body top quark FCNC decays $t\to c V$ at one-loop level induced by the scalar LQ $S_1$ in Section \ref{sec-one-loop}. For the dipole current we have:
\begin{eqnarray}
%f^{\gamma}_{VL} &=& 0\\
%f^{\gamma}_{VR} &=& 0\\
f^{g}_{TL} &=& g_{1 L}^{33} g_{1 R}^{23*} \frac{1  }{16 \pi ^2} g_s m_{\tau } \times  C_2(0,m_t^2,m_c^2,M_{S_1}^2,M_{S_1}^2,m_{\tau }^2), \label{eq:fVtcgFULL}\\
f^{\gamma}_{TL} &=& -g_{1 L}^{33} g_{1 R}^{23*} \frac{1  }{48 \pi ^2} e m_{\tau }\nonumber \\
& &\times\Big(3
    C_1(m_c^2,0,m_t^2,M_{S_1}^2,m_{\tau }^2,m_{\tau
   }^2)+ C_2(0,m_t^2,m_c^2,M_{S_1}^2,M_{S_1}^2,m_{\tau }^2) \nonumber \\
   & &+3
    C_2(m_c^2,0,m_t^2,M_{S_1}^2,m_{\tau }^2,m_{\tau }^2) \Big), \label{eq:fVtcAFULL}\\
f^{Z}_{TL} &=& g_{1 L}^{33} g_{1 R}^{23*}  \frac{1}{96 \pi ^2} \frac{e}{ c_W s_W } m_{\tau } \nonumber \\
& & \times\Big(3 (s_W^2-c_W^2)  C_1(m_c^2,m_Z^2,m_t^2,M_{S_1}^2,m_{\tau }^2,m_{\tau }^2)\nonumber \\
& &+2 s_W^2 \big(3  C_2(m_c^2,m_Z^2,m_t^2,M_{S_1}^2,m_{\tau }^2,m_{\tau }^2)+ C_2(m_Z^2,m_t^2,m_c^2,M_{S_1}^2,M_{S_1}^2,m_{\tau }^2)\big)\Big),\\
f^{g}_{TR} &=& f^{\gamma}_{TR}  = f^{Z}_{TR} = 0.
\end{eqnarray} 
Note that the absence of right-handed dipole current is because of the coupling textures we considered in Table.\ref{Table:parameters}.

For the monopole current which appears for the massive gauge boson $Z$, we have:
\begin{eqnarray}
%%%%%%
f^{Z}_{VL} &=& g_{1 L}^{33} g_{1 R}^{23*}  \frac{1}{96 \pi ^2} \frac{e}{ c_W s_W }   \frac{m_{\tau } m_c}{m_c^2-m_t^2} \nonumber \\
   & & \times\Big((3 c_W^2-s_W^2) \big( B_0(m_t^2,m_{\tau }^2,M_{S_1}^2)- B_0(m_c^2,m_{\tau
   }^2,M_{S_1}^2)\big) \nonumber \\
   & & +2 s_W^2 (m_c^2-m_t^2) \big(3  C_0(m_c^2,m_Z^2,m_t^2,M_{S_1}^2,m_{\tau }^2,m_{\tau }^2)+3  C_1(m_c^2,m_Z^2,m_t^2,M_{S_1}^2,m_{\tau }^2,m_{\tau
   }^2) \nonumber \\
   & & +3  C_2(m_c^2,m_Z^2,m_t^2,M_{S_1}^2,m_{\tau }^2,m_{\tau }^2)+ C_2(m_Z^2,m_t^2,m_c^2,M_{S_1}^2,M_{S_1}^2,m_{\tau }^2)\big)\Big),\\
f^{Z}_{VR} &=& g_{1 L}^{33} g_{1 R}^{23*}  \frac{1}{96 \pi ^2} \frac{e}{ c_W s_W } \frac{m_{\tau } m_t}{m_c^2-m_t^2} \nonumber \\
& &\times\bigg(4 s_W^2 \big( B_0(m_c^2,m_{\tau }^2,M_{S_1}^2)-  B_0(m_t^2,m_{\tau }^2,M_{S_1}^2)\big)\nonumber \\
& &+(m_c^2-m_t^2) \Big(2 s_W^2  C_2(m_Z^2,m_t^2,m_c^2,M_{S_1}^2,M_{S_1}^2,m_{\tau }^2) -3 (c_W^2-s_W^2) \big( C_0(m_c^2,m_Z^2,m_t^2,M_{S_1}^2,m_{\tau }^2,m_{\tau }^2)\nonumber \\
& &+ C_1(m_c^2,m_Z^2,m_t^2,M_{S_1}^2,m_{\tau }^2,m_{\tau }^2)+ C_2(m_c^2,m_Z^2,m_t^2,M_{S_1}^2,m_{\tau }^2,m_{\tau }^2)\big)\Big)\bigg).
\end{eqnarray} 

Coefficients $B_i, C_i$ in the above are defined in the general one-loop tensor integral \cite{Hahn:1998yk,Denner:1991kt}:
\begin{eqnarray}
T^{N}_{\mu_{1}\ldots\mu_{P}} (p_{1},\ldots,p_{N-1},m_{0},\ldots,m_{N-1})=
\frac{(2\pi\mu)^{4-D}}{i\pi^{2}}\int d^{D}\!q
\frac{q_{\mu _{1}}\cdots q_{\mu _{P}}}{D_{0}D_{1}\cdots D_{N-1}},
\end{eqnarray} 
with the following denominators:
\begin{eqnarray}
D_{0}= q^{2}-m_{0}^{2}+i\varepsilon, \qquad
D_{i}= (q+p_{i})^{2}-m_{i}^{2}+i\varepsilon, \qquad i=1,\ldots,N-1.
\end{eqnarray}
Then one can perform the decompositions as follows:
\begin{eqnarray}
B_{\mu } & = & p_{1\mu } B_{1}, \\
C_{\mu } & = & p_{1\mu } C_{1} + p_{2\mu } C_{2}.
\end{eqnarray} 
To carry out the numerical calculations with the public package LoopTools, one needs to impose the parameter conventions of LoopTools by the following transformation:
\begin{eqnarray}
B_{i} (p_{1}, m_{0},m_1) & \to & B_{i}(p_{1}^2, m_0^2, m_1^2), \\
C_i (p_{1}, p_2, m_{0}, m_1, m_2) & \to & C_{i}\big(p_{1}^2, (p_1-p_2)^2, p_2^2, m_0^2, m_1^2, m_2^2\big).
\end{eqnarray} 
Note that the loop function coefficients presented in this Appendix has been applied with the transformation.

\begin{figure}[ht]
\begin{center}
\includegraphics[width=7cm]{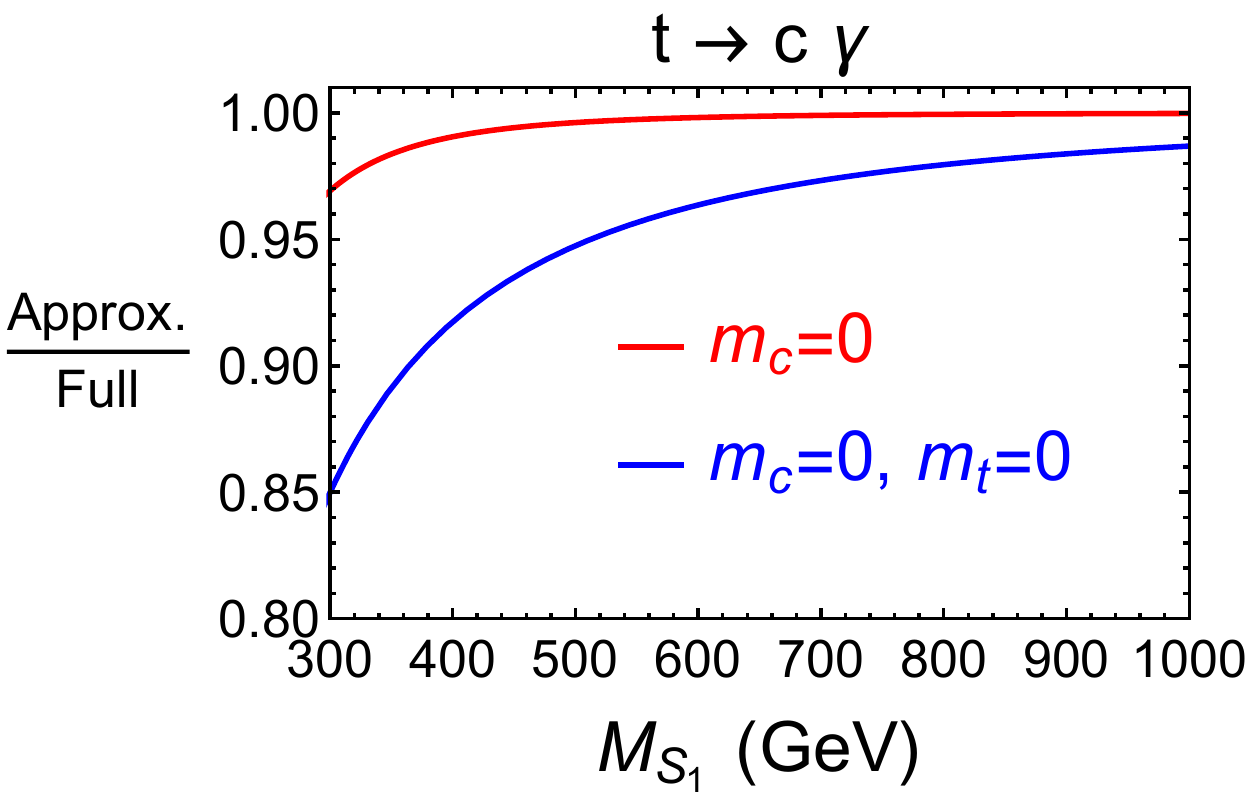}
\includegraphics[width=7cm]{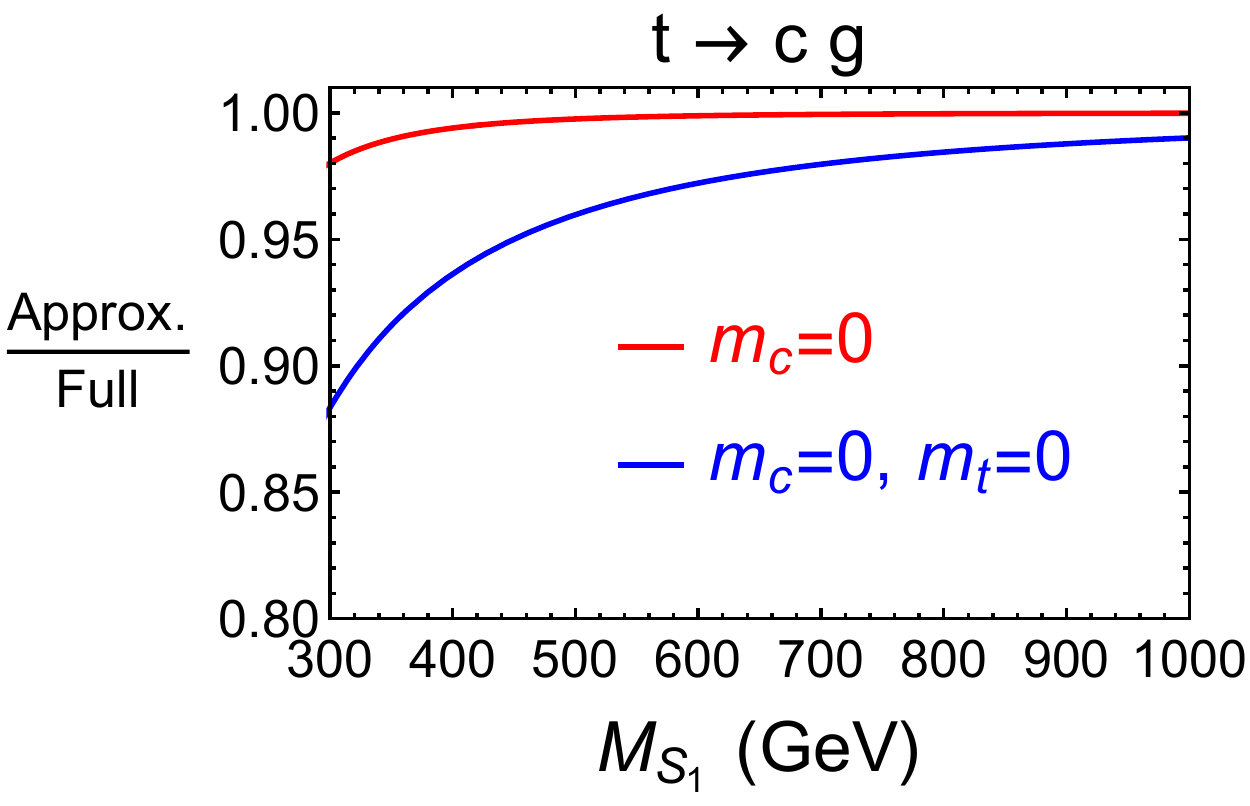}
\caption{The comparison of the approximations in Eq.(\ref{eq:fVtcg}) and Eq.(\ref{eq:fVtcA}) to the full results in Eq.(\ref{eq:fVtcgFULL}) and Eq.(\ref{eq:fVtcAFULL}).}
\label{fig:loop-comparison}
\end{center}
\end{figure}

In all of the numerical results presented in this work, we have used the complete expressions with loop functions. However, for $M_{S_1}\simeq \mathcal{O}(1)\, {\rm TeV}$ which indicates $M_{S_1}\gg m_t,m_c,m_\tau$, one can have the following compact approximations by setting $m_c=0$ for the massless cases of photon and gluon, 
\begin{eqnarray}
f^{g}_{TL} &\simeq& \frac{1  }{16 \pi ^2} g_s m_{\tau }  \frac{g_{1 L}^{33} g_{1 R}^{23*} }{M_{S_1}^2} \nonumber \\
& &\times \frac{1}{12} \big(  -6 \left( 22 x_{\tau } x_t +3 x_t+16 x_{\tau }+6\right) \log x_{\tau }-49 x_t-48 \big),\\
f^{\gamma}_{TL} &\simeq& - \frac{1  }{16 \pi ^2} e m_{\tau }  \frac{g_{1 L}^{33} g_{1 R}^{23*} }{M_{S_1}^2} \nonumber \\
& &\times \frac{1}{3}\big( \frac{1}{6} \left(14 x_t x_{\tau }+x_t+9 x_{\tau }+3\right)+\left(x_t+1\right) x_{\tau } \log x_{\tau }\big),\\
f^{g}_{TR} &=& f^{\gamma}_{TR}  = f^{Z}_{TR} = 0,
\end{eqnarray} 
where $x_\tau=m_\tau^2/M_{S_1}^2, x_t=m_t^2/M_{S_1}^2$. Note that we have kept the first order effect of top quark mass $m_t$ in the expansion due to its relatively large value. To illustrate this, in Fig.\ref{fig:loop-comparison} we show the comparison of the approximations to the full results. In the cases of massive $Z$ boson with two external massive particles in the 3-point loop functions, even the approximated expressions are tediously long \cite{Denner:1991kt} and we do not proceed with the reductions but keeping the full expression.

\bibliographystyle{JHEP}
\bibliography{B-v4}

\end{document}